\documentclass[12pt]{article}

\usepackage{epsfig,epstopdf,latexsym,amsfonts,amsmath,amsthm,amssymb,amsbsy,multirow,slashed,wasysym,textcomp,wrapfig,graphicx,psfrag,booktabs,bbm,comment}

\usepackage[toc]{appendix}

\usepackage{color}
\usepackage{datetime}
\usepackage[
      colorlinks=false,
      linkcolor=darkblue,  
      urlcolor=blue,    
      filecolor=blue,     
      citecolor=red,
linktocpage=true,
      pdfstartview=FitV,
      bookmarksopen=true    
      ]{hyperref}

\ifpdf
\fi

\numberwithin{equation}{section}

\renewcommand{\theequation}{\arabic{section}.\arabic{equation}}

\def\coeff#1#2{\relax{\textstyle {#1 \over #2}}\displaystyle}

\def\IC{\mathbb{C}}

\def\IR{\mathbb{R}}
\def\ZZ{\mathbb{Z}}

\def\cB{{\cal B}}

\def\cF{{\cal F}}

\def\cL{{\cal L}}

\def\cN{{\cal N}}
\def\cO{{\cal O}}

\def\cQ{{\cal Q}}

\def\Neql#1{{\cal N}\!=\!{#1}}

\definecolor{cardinal}{rgb}{0.6,0,0}
\definecolor{darkgreen}{rgb}{0,0.5,0}
\definecolor{golden}{rgb}{0.92, 0.7, 0}
\definecolor{midnight}{rgb}{0, 0, 0.5}
\definecolor{darkblue}{rgb}{0.2, 0, 0.8}

\begin{document}

\begin{titlepage}



\bigskip
\bigskip

\centerline{\Large \bf Doubly-Fluctuating BPS Solutions in Six Dimensions}

\medskip
\bigskip
\bigskip
\centerline{{\bf Benjamin E. Niehoff$^1$  and Nicholas P. Warner$^{1,2,3}$}}
\centerline{{\bf }}
\bigskip
\centerline{$^1$ Department of Physics and Astronomy}
\centerline{University of Southern California} \centerline{Los
Angeles, CA 90089, USA}
\bigskip
\centerline{$^2$ Institut de Physique Th\'eorique, }
\centerline{CEA Saclay, CNRS-URA 2306, 91191 Gif sur Yvette, France}
\bigskip
\centerline{$^3$ Institut des Hautes Etudes Scientifiques}
\centerline{ Le Bois-Marie, 35 route de Chartres}
\centerline{Bures-sur-Yvette, 91440, France}
\bigskip
\bigskip
\centerline{{\rm bniehoff@usc.edu, ~warner@usc.edu} }
\bigskip
\bigskip

\begin{abstract}
\noindent We analyze the BPS solutions of minimal supergravity coupled to an anti-self-dual tensor multiplet  in six dimensions and find solutions that fluctuate non-trivially as a function of two variables.  We consider families of solutions coming from KKM monopoles fibered over Gibbons-Hawking metrics or, equivalently, non-trivial $T^2$ fibrations over an $\IR^3$ base.  We find smooth microstate geometries that depend upon many functions of one variable, but each such function depends upon a different direction inside the   $T^2$  so that the complete solution depends non-trivially upon the whole  $T^2$.  We comment on the implications of our results for the construction of a general superstratum.

\end{abstract}

\end{titlepage}

\tableofcontents
\newpage
\section{Introduction}

The general classification of BPS solutions in supergravity is expected to have a wide range of applications, ranging from holography to the description of black-hole microstates.  This issue has become particularly significant in six dimensions for several reasons.  First, it is perhaps the simplest setting of the D1-D5-P system, which lies at the heart of the stringy description of BPS black holes with macroscopic horizons \cite{Strominger:1996sh}  and the possible construction of microstate geometries (see \cite{Lunin:2001jy, Lunin:2002iz}, and for reviews, see  \cite{Mathur:2005zp, Bena:2007kg, Balasubramanian:2008da,Skenderis:2008qn, Chowdhury:2010ct}). Secondly,  six-dimensional supergravity underlies the study of $AdS_3 \times S^3$ holography (see, for example,  \cite{Seiberg:1999xz,Arutyunov:2000by,Balasubramanian:2000rt, Lunin:2002bj,Kanitscheider:2006zf}).  Thirdly, it has become evident that while five-dimensional microstate geometries can resolve black-hole singularities and provide rich families of solutions that sample the typical sector of the black-hole conformal field theory, there are not enough such microstate geometries  to sample the states of the black hole with sufficient density so as to yield a semi-classical description of the thermodynamics \cite{Bena:2006is,Bena:2007qc,deBoer:2008zn,deBoer:2009un}.  The five-dimensional microstate geometries are trivial compactifications of IIB supergravity and M-theory and it is hoped that the incorporation of fluctuations in six, or more, dimensions will greatly extend the phase-space coverage of the microstate geometries.   Finally, there was something of a breakthrough in the analysis of six-dimensional supergravity in that  the BPS equations of the simplest, but probably most important class of such supergravities are substantially linear \cite{Bena:2011dd}.  This raises the possibility of finding new classes and families of solution and analyzing the phase space structure more completely.

The jump from the five-dimensional to the six-dimensional BPS system is not expected to be merely incremental in terms of solutions and structure.  While five-dimensions is just enough to resolve black-hole singularities, one of the key messages in \cite{Bena:2006is,Bena:2007qc,deBoer:2008zn} is that  this formulation is still too rigid.  
The index computations of \cite{deBoer:2008zn,deBoer:2009un} show that fluctuations around five-dimensional backgrounds involving a simple ``graviton gas''  can produce a denser but still inadequate (for semi-classical thermodynamics) sampling of microstates.  One way to evade this conclusion is to put fluctuations on non-perturbative structures in higher dimensions and for this even the humble, fluctuating supertube could, in principle, be sufficient particularly if the entropy enhancement mechanism  \cite{Bena:2008nh,Bena:2010gg} can be fully realized.  A somewhat more radical proposal was made in \cite{Bena:2011uw}, where it was proposed that there should be a new kind of solitonic object in six dimensions: the {\it superstratum}.  This was conjectured to be a smooth, six-dimensional microstate geometry whose shape and density modes can be general functions of two variables.  These objects also carry three charges and two independent dipole charges and are thus very natural, fundamental constituents of the three-charge black hole.  The construction of superstrata, even in a restricted form, is one of the major motivations for this paper.

The theory that underlies much of the work on five-dimensional microstate geometries is $\cN =2$ supergravity coupled to two vector multiplets in five dimensions.  The three vector fields (including the graviphoton) are sourced by the essential three charges of the system.  This theory, when uplifted becomes 
 minimal supergravity coupled to an anti-self-dual tensor multiplet  in six dimensions.  One may also think of this in terms of a compactification of the D1-D5-P system in IIB supergravity on a four-torus. The BPS equations of this system were first extracted in   \cite{Gutowski:2003rg,Cariglia:2004kk} but the form seemed hopelessly non-linear at all levels.  In spite of this, one could construct a limited set of non-trivial solutions using spectral flow techniques \cite{Lunin:2004uu, Ford:2006yb}.  However, it was recently shown in \cite{Bena:2011dd} that while the requirements on the five-dimensional spatial background is essentially non-linear, the BPS equations that determine all the charges, two sets of the magnetic fluxes and the angular momentum are, in fact, linear.  This means that there are certainly interesting new classes of BPS solution within reach and some of these have already been obtained  \cite{Bena:2011dd,Niehoff:2012wu,Bobev:2012af,Vasilakis:2013tjs}. These solutions typically start with a simple background geometry which is then decorated with non-trivial charges, magnetic fluxes and sometimes shape modes.  Such solutions, while interesting, are usually singular.  
 
 Our purpose here is to start with non-trivial geometric backgrounds involving Kaluza-Klein monopoles (KKM's), analyze the BPS equations and then find new, smooth microstate geometries that fluctuate non-trivially as a function of two variables.  In our analysis of the BPS equations, we  find features that fit very well with the constructive algorithm outlined in \cite{Bena:2011uw}: We see that the BPS system can accommodate the tilting and boosting  of the parallel D1 and D5 branes, and the addition of momentum and angular momentum densities,  in such a manner that one induces d1-d5 dipole densities by reorienting the  D1-D5 charge densities and that all of this can be achieved in a way that makes the densities into functions of two variables.  It was further argued in \cite{Bena:2011uw} that such solutions can be made smooth via the addition of an appropriately varying KKM configuration.  This last step generically involves solving non-linear equations and, to date, the only known non-trivial solutions come from freezing fluctuations of the KKM configuration\footnote{There have  been some interesting recent attempts to use string amplitude calculations to determine at least the perturbative form of generic fluctuating KKM's \cite{Giusto:2011fy, Giusto:2009qq, Giusto:2012gt, Giusto:2012jx}.}, which reduces the problem to a linear system.   We use this strategy here and freeze the KKM configuration.  In this sense, our new solutions  may be thought of as a form of semi-rigid superstrata in that the KKM's are rigid but the charge densities fluctuate.  As we will see, the fluctuations are limited by the rigidity of the Kaluza-Klein monopoles and are ultimately sourced by an arbitrary, but finite, number of functions of one variable.  One the other hand, the one variable that appears in each such source can be a different linear combination of the two variables that parametrize the fluctuations and so our solutions are indeed doubly-fluctuating.

 In Section 2 we review the BPS equations of minimal supergravity coupled to an anti-self-dual tensor multiplet  in six dimensions. In Section 3 we specialize to a background that involves a collection of KKM's fibered over a four-dimensional Gibbons-Hawking (GH) base manifold.  This is then recast as a torus fibration over a flat, $\IR^3$ base with the circles of the torus fibered non-trivially as a set of doubled, independent  KKM's.  The general, doubly fluctuating solution will depend non-trivially on both directions on the torus. In Section 4 we reduce the BPS equations on this double KKM fibration and find a large class of  fluctuating solutions that are governed by a single differential operator.  This operator is particularly interesting because it is the six-dimensional Laplacian reduced on the ``time coordinate.''   The appearance of such an operator is not  surprising because BPS solutions necessarily have some form of time-translation invariance, but  the new feature here is that the form of the six-dimensional supersymmetry means that this invariance is actually a null translation and the reduction of the Laplacian thus produces a degenerate operator.  In Section 5 we examine the regularity of such solutions, first by working with a particularly simple example and then using this to infer the structure of the general solutions.  We find that the rigidity of the doubled set of KKM monopoles restricts the fluctuations to be sourced by functions of one variable but that the particular variable in the density functions can slice the torus in different ways depending upon the KKM charges.  Section 6 contains our conclusions and some technical generalizations of our BPS analysis appears in an appendix.

\section{The BPS solutions in six dimensions}
\label{SixDBPS}

The six-dimensional system we study is $\Neql 1$ minimal supergravity coupled to one anti-self-dual tensor multiplet. This theory, upon trivial dimensional reduction, gives rise to $\Neql 2$, five-dimensional  supergravity coupled to two vector multiplets and thus contains three independent electromagnetic fields.  In the six-dimensional theory, the graviton multiplet contains a self-dual tensor field and so the entire bosonic sector consists of the graviton, the dilaton and an unconstrained $2$-form gauge field with a $3$-form field strength.  The BPS equations of this theory were constructed in  \cite{Gutowski:2003rg,Cariglia:2004kk} and in \cite{Bena:2011dd} it was shown that the BPS system could be dramatically simplified and that most of the equations could be reduced to a linear system. We now review this structure in some detail.

\subsection{The background geometry}
\label{Sect:Background}
 
For supersymmetric solutions the metric is necessarily highly constrained \cite{Gutowski:2003rg}: 
\begin{equation}
ds_6^2 ~=~   2 H^{-1} (dv+\beta) \big(du +  k  ~+~ \coeff{1}{2}\,  \cF\, (dv+\beta)\big) ~-~  H \, ds_4^2(\cB)\,.   \label{sixmet}
\end{equation}
where the  metric on the four-dimensional base,  $\cB$, is written in terms of components as:
\begin{equation}
 ds_4^2~=~  h_{\mu \nu} dx^\mu dx^\nu \,. 
 \end{equation}
Supersymmetry necessarily implies a time-translation invariance and in this setting it means that the entire six-dimensional background must independent of $u$.  We are interchanging the labeling of the coordinates $u,v$ relative to \cite{Gutowski:2003rg,Cariglia:2004kk} but we will otherwise follow the conventions of \cite{Gutowski:2003rg}.   We start by  introducing the frames, $\{e^+,e^-,e^a\}$ defined by: 
\begin{equation} 
e^+ \equiv H^{-1} \big(dv + \beta \big)  \,,  \qquad
e^- ~\equiv~  du +k    + \coeff{1}{2} \,{\cal F} H \, e^+  \,, \qquad 
e^a = H^{1 \over 2} \,  \hat{e}^a{}_\mu \, dx^\mu \,,  \label{eqn:frames}
\end{equation} 
so that we have: 
\begin{equation} 
ds^2 ~=~   2 e^+ e^- ~-~ \delta_{ab}\, e^a  \, e^b \,. 
\end{equation} 

In $d$ dimensions, again following \cite{Gutowski:2003rg}, we define the Hodge star $*_d$ to act on a $p$-form via
\begin{equation}
 *_d\, (dx^{m_1}\wedge\cdots\wedge dx^{m_p})
 ~=~  {1\over (d-p)!} \, dx^{n_1}\wedge\cdots\wedge dx^{n_{d-p}}\,  \epsilon_{n_1\dots n_{d-p}}{}^{m_1\dots m_p} \,. \label{stupiddual}
\end{equation}
where we will use the orientations:
\begin{equation} 
\epsilon^{+-1234}  ~=~ \epsilon^{1234} ~=~ +1 \,.
\end{equation} 
Note that (\ref{stupiddual}) is not the typical convention (like that of \cite{Nakahara:2003nw})  for the Hodge dual.

The first step in simplifying the BPS equations is to refer everything to geometry on the $U(1)$ fiber described by $v$ over the base manifold, $\cB$, with coordinates $x^\mu$.  To that end, we introduce the restricted exterior derivative, ${\tilde{d}}$, acting on a $p$-form,   $\Phi \in \Lambda^p ({\cal B})$, by:
\begin{eqnarray} 
 \Phi &=& {1 \over p!} \, \Phi_{\mu_1 \dots \mu_p} (x,v) \, dx^{\mu_1} \wedge \ldots \wedge  dx^{\mu_p} \,, \\
 {\tilde{d}} \Phi &\equiv&  {1 \over (p+1)!} \, (p+1) \, {\partial \over \partial x^{[\nu}} \Phi_{\mu_1 \dots \mu_p]} \,  dx^\nu \wedge  dx^{\mu_1}
 \wedge \ldots \wedge dx^{\mu_p}\,.
\end{eqnarray} 
and we define a Kaluza-Klein covariant differential operator, $D$,  by:
\begin{equation} 
D \Phi ~\equiv~ {\tilde{d}} \Phi ~-~ \beta \wedge {\dot{\Phi}},  \label{Ddefn}
\end{equation} 
where we denote a derivative with respect to $v$ by a dot.  

Define the field strength, $\Theta_3$, by 
\begin{equation} 
\Theta_3  \equiv  D \beta \,,  \label{Theta3defn}
\end{equation} 
then supersymmetry requires this to be self-dual: 
\begin{equation} 
\Theta_3  ~\equiv~    *_4  \Theta_3 \,. \label{betacond}
\end{equation} 
The base, $\cB$,  is required to be ``almost hyper-K\"ahler,''  or, more precisely, it is hyper-complex in that there are three anti-self-dual $2$-forms: 
\begin{equation}
 J^{(A)}  ~\equiv~ \coeff{1}{2}\,   {J^{(A)}}_{mn}  \, dx^m  \wedge dx^ n \label{Jdefn}  \,,
 \end{equation}
that satisfy the quaternionic algebra: 
\begin{equation}
{J^{(A)}}{}^m{}_p {J^{(B)}}{}^p{}_n = \epsilon^{ABC}\,{J^{(C)}}{}^m{}_n ~-~  \delta^ {AB} \,\delta^m_n \,.
\label{Jalg} 
 \end{equation}
These forms are also required to satisfy the differential identity:
\begin{equation}
\tilde d J^{(A)}  ~=~    \partial_v \big (\beta \wedge J^{(A)})  \label{Jcond} \,,
 \end{equation}
where $\partial_v\Phi$ denotes the Lie derivative of a quantity $\Phi$ with respect to the tangent vector $\partial \over \partial v$.   The self-duality condition (\ref{betacond}) guarantees the integrability of  (\ref{Jcond}).   Note that we are using the conventions of \cite{Gutowski:2003rg} in the definition of the $J^{(A)}$ and that one must make the replacement $J^{(A)} \to - J^{(A)}$ in order to go to the conventions of \cite{Cariglia:2004kk}.

Finally, it is convenient to introduce the the anti-self-dual $2$-forms, $\psi$ and $\hat \psi$ defined by:
\begin{equation}
\psi  ~\equiv~ H \, \hat \psi  ~\equiv~  \coeff{1}{16} \, H\,  \epsilon^{ABC}  \, J^{(A)}{}^{mn} \dot {J}^{(B)}{}_{mn} \, J^{(C)} \label{psidefn}  \,,
\end{equation}
%

\subsection{The tensor gauge field and the dilation}
\label{Sect:OtherFields}

The three-form tensor gauge field satisfies a Bianchi identity  and has the equation of motion:
\begin{equation}
d\, G~=~  0\,, \qquad d \big(e^{2 \sqrt{2} \phi}\, *_6 G \big) ~=~  0  \label{Geqna}   \,,
\end{equation}
However, supersymmetry imposes  strong constraints on the form of $G$ and these constraints can be significantly simplified 
by writing $G$ and  its dual  in terms of electric and magnetic parts \cite{Bena:2011dd}:
\begin{eqnarray}  
G  &=&  d \big[ - \coeff{1}{2}\,Z_1^{-1}\,(du + k ) \wedge (dv + \beta)\, \big] ~+~ \widehat G_1 \,,  \label{niceGform}   \\
e^{2\, \sqrt{2} \phi} \,*_6 G  &=&   d \big[ - \coeff{1}{2}\,Z_2^{-1}\,(du + k ) \wedge (dv + \beta)\,   \big] ~+~ \widehat G_2  \,,\label{nicedualGform}   
\end{eqnarray}
where 
\begin{eqnarray}  
 \widehat G_1  &\equiv&   \coeff{1}{2} *_4 (D Z_2 + \dot{\beta} Z_2) ~+~ \coeff{1}{2}\,  (dv+ \beta) \wedge \Theta_1 \,,  \label{G1hat}   \\
 \widehat G_2 &\equiv&   \coeff{1}{2} *_4 (D Z_1 + \dot{\beta} Z_1) ~+~   \coeff{1}{2}\,  (dv+ \beta) \wedge \Theta_2\,. \label{G2hat}   
\end{eqnarray}
Thus the flux is  defined in terms of two electrostatic potentials, $Z_j$, and two two-forms,  $\Theta_j$.  Note that we have rescaled $\Theta_j \to \frac{1}{2} \Theta_j$ relative to the conventions of \cite{Bena:2011dd}.  This new choice of normalization is probably the simplest way to map the six-dimensional BPS equations onto the standard form of the five-dimensional BPS equations.  The Bianchi identities and Maxwell equations (\ref{Geqna})  require the closure of the $\widehat G_j$, which shows that these quantities do indeed measure a conserved magnetic charge.  

The two-form fluxes are required to be self-dual up to shifts by the two-form, $\psi$:
\begin{eqnarray}
*_4 \Theta_1   &=& \Theta_1  -   4 \,  e^{-\sqrt{2}\phi}\, \psi ~=~ \Theta_1  -  4 \,  Z_2\, \hat \psi  \,,  \label{Theta1dual} \\
*_4 \Theta_2   &=& \Theta_2  -   4 \,  e^{\sqrt{2}\phi}\, \psi ~=~ \Theta_2  -  4 \,  Z_1\, \hat \psi  \,, \label{Theta2dual}
\end{eqnarray}
and so if one defines 
\begin{equation}
\widehat \Theta_1  ~\equiv~  \Theta_1 ~-~ 2 \,  Z_2\, \hat \psi \,, \qquad \widehat  \Theta_2 ~\equiv~  \Theta_2 ~-~   2 \,  Z_1\, \hat \psi  \,,   \label{modTheta}
\end{equation}
then these two-forms are self-dual:
\begin{equation} 
\widehat \Theta_j  ~\equiv~    *_4  \widehat \Theta_j   \,, \qquad j=1,2 \,.   \label{modThetaSD}
\end{equation}

Supersymmetry also requires that the electric potentials be related to the warp factor and dilaton in a simple generalization of the ``floating brane Ansatz'' \cite{Bena:2009fi}:
\begin{equation}
Z_1 ~\equiv~H\, e^{\sqrt{2}\phi}\,,\qquad Z_2 ~\equiv~ H\, e^{-\sqrt{2}\phi}\,.  \label{ZDefns}
\end{equation}

The form of $G$ required by supersymmetry makes  its self-dual part (in six dimensions)  the same as the self-dual part  of spin connection.  This means that the supersymmetry variations become trivial and that  the supersymmetry parameters are constant in the frames and coordinates introduced above:
\begin{equation} 
\partial_M \epsilon ~=~ 0  \,. \label{consteps}
\end{equation} 
%

\subsection{The angular momentum vector and the momentum potential}
\label{Sect:Z3k}

It is convenient to define:
\begin{equation}
L ~\equiv~  \dot{k }~+~ \coeff{1}{2}\, \cF \dot \beta ~-~  \coeff{1}{2}\,  D \cF ~=~  \coeff{1}{2}\,  D Z_3 ~-~ \coeff{1}{2}\, \dot \beta  \, Z_3 ~+~  \dot{k }  \,,   \label{Ldefn}
\end{equation}
where we have introduced the momentum potential, 
\begin{equation}
Z_3  ~\equiv~  - \cF    \,,   \label{Z3defn}
\end{equation}
so as to make direct contact with the five-dimensional formulation \cite{Bena:2005va,Berglund:2005vb,Bena:2007kg}.   

The quantity, $L$,   is gauge invariant under the transformation:
\begin{equation}
 \cF ~\to~  \cF +2\, \partial_v f   \,, \qquad  k  ~\to~ k  + D f  \,,   \label{gaugetrf}
\end{equation}
for any function, $f(v, x^m)$.  This transformation is induced by a coordinate change $u \to u + f(v, x^m)$ in the metric (\ref{sixmet}).

\subsection{The BPS equations}

Once one has constructed a background that satisfies the conditions stipulated in Section \ref{Sect:Background} then the remaining BPS equations are linear.
First, the electrostatic potentials are related to the magnetic fluxes via:
\begin{equation}
\tilde d \Theta_2 ~=~  \partial_v  \big[*_4 (D Z_1 + \dot{\beta} Z_1)~+~  \beta \wedge \Theta_2 \big] \label{Theta2eqna}\,,
\end{equation}
\begin{equation}
D *_4 (D Z_1 + \dot{\beta} Z_1) = -  \Theta_2 \wedge  D \beta\,, \label{Z1eqn}
\end{equation}
and
\begin{equation}
\tilde d \Theta_1~=~  \partial_v  \big[*_4 (D Z_2 + \dot{\beta} Z_2)~+~  \beta\wedge \Theta_1 \big] \label{Theta1eqna} \,,
\end{equation}
\begin{equation}
D *_4 (D Z_2 + \dot{\beta} Z_2) = -  \,\Theta_1 \wedge  D \beta\,. \label{Z2eqn}
\end{equation}
These constitute  linear systems in $(Z_1,\Theta_2)$ and $(Z_2,\Theta_1)$ independently.

The last layer of the BPS equations then relate the angular momentum vector to the momentum potential, $Z_3 = - \cF$:
\begin{eqnarray}
*_4 D *_4 L &=& \coeff{1}{2}\, H h^{mn}\partial_v^2 (H h_{mn}) + \coeff{1}{4}\, \partial_v (H h^{mn}) \,\partial_v (H h_{mn}) - 2\, \dot{\beta}_m\, L^m + 2\, H^2 \,\dot{\phi}^2  \nonumber\\
&& -  *_4 \Big[\, \coeff{1}{2}\, \Theta_1 \wedge \Theta_2 ~-~  H^{-1}  \psi  \wedge D k  \, \Big] \,.  \label{Feqn}
\end{eqnarray}
\begin{eqnarray}
D k  ~+~  *_4 D k   &=&     Z_1\, \Theta_1 +   Z_2\, \Theta_2  ~-~  \cF \, D\beta - 4\, H\, \psi \nonumber \\
&=&     Z_1\, \big (\Theta_1 -2 \, Z_2\, \hat \psi\big) +   Z_2\, \big(\Theta_2 - 2\,Z_1\, \hat \psi\big) ~-~  \cF \, D\beta    \nonumber \\
&=&     Z_1\,   \widehat \Theta_1 ~+~    Z_2\, \widehat \Theta_2   ~+~ Z_3 \, \Theta_3     \label{angmomeqn} \,.
\end{eqnarray}
Once again, this is a linear system for $(Z_3,k )$.

Note that if the background and fields are all $v$-independent, then these BPS equations reduce to:
\begin{equation}
\tilde d \Theta_1 ~=~ \tilde d \Theta_1  ~=~ 0\,, \qquad \Theta_j ~=~ *_4 \Theta_j  \,, \label{Theta5d}
\end{equation}
\begin{eqnarray}
\nabla^2_{(4)} Z_1 &=& - *_4  \tilde d *_4  \tilde d Z_1  ~ = ~     *_4  \Theta_2 \wedge   \Theta_3  \,, \label{BPSfiveD1} \\
  \nabla^2_{(4)} Z_2 &=& -*_4   \tilde d *_4  \tilde d Z_2  ~ = ~   *_4   \Theta_1 \wedge   \Theta_3 \,,  \label{BPSfiveD2} \\
 \nabla^2_{(4)} Z_3 &=&- *_4 \tilde d *_4 \tilde d Z_3 ~=~ - 2\, *_4 \tilde d *_4 L ~=~   *_4   \Theta_1 \wedge \Theta_2  \,, \label{BPSfiveD3}
\end{eqnarray}
and finally
\begin{equation}
\tilde d k  ~+~  *_4 \tilde d  k   ~=~  \sum_{I=1}^3 ~ Z_I  \Theta_I  \,.  \label{BPSfiveD4}
\end{equation}
These are, of course, the canonically normalized five-dimensional BPS equations  \cite{Bena:2005va,Berglund:2005vb,Bena:2007kg}.

\section{BPS solutions with a Gibbons-Hawking base}
\label{Sect:GHbase}

We now simplify the BPS system by considering background geometries that are completely independent of $v$ and in which one has a generic, multi centered Gibbons-Hawking (GH) metric on the base. We will also assume that the vector field, $\beta$, defining the fibration is also $v$ independent.  While we will simplify the base and the fiber potential in this manner, we will allow the fluxes, warp factors and dilaton to be $v$ dependent, 

\subsection{The background geometry}
\label{Sect:GHbackground}

We start by taking
\begin{equation}
ds_4^2 ~=~ V^{-1} \, \big( d\psi + A)^2  ~+~ V\, d \vec y \cdot d
\vec y \,, \label{GHmetric}
\end{equation}
where, on the flat $\IR^3$ defined by the coordinates $\vec y$, one has:
\begin{equation}
\nabla^2 V ~=~ 0\,, \qquad \vec \nabla \times \vec A ~=~ \vec \nabla V\,.
\label{AVreln}
\end{equation}
We take $V$ to have the form
\begin{equation}
 V ~=~ h ~+~ \sum_{j=1}^N \, \frac{q_j}{|\vec y - \vec y^{(j)}| } \,, 
\label{Vform}
\end{equation}
for some fixed points, $\vec y^{(j)} \in \IR^3$ and some charges, $q_j \in \ZZ$.

We use the following set of frames:
\begin{equation}
\hat e^1~=~ V^{-{1\over 2}}\, (d\psi ~+~ A) \,,
\qquad \hat e^{a+1} ~=~ V^{1\over 2}\, dy^a \,, \quad a=1,2,3 \,.
\label{GHframes}
\end{equation}
and define two associated sets of two-forms:
\begin{equation}
\Omega_\pm^{(a)} ~\equiv~ \hat e^1  \wedge \hat e^{a+1} ~\pm~ \coeff{1}{2}\, \epsilon_{abc}\,\hat e^{b+1}  \wedge \hat e^{c+1} \,, \qquad a =1,2,3\,.\
\label{twoforms}
\end{equation}
The two-forms, $\Omega_-^{(a)}$, are anti-self-dual,  harmonic and non-normalizable  and they define the hyper-K\"ahler  structure on the base.  We therefore identify the $J^{(A)}$ in (\ref{Jdefn}) with the $\Omega_-^{(A)}$.  These are manifestly $v$-independent and thus, from  (\ref{psidefn}), we have $\psi \equiv 0$. Moreover, $\Theta_3$ must be harmonic.

The forms, $\Omega_+^{(a)}$, are self-dual and can be used to construct harmonic fluxes that are dual to the two-cycles.  In particular, we will lake 
\begin{equation}
\Theta_3 ~ \equiv~ - \sum_{a=1}^3 \, \big(\partial_a \big( V^{-1}\, K_3 \big)\big) \,
\Omega_+^{(a)} \,.
\label{harmtwoform}
\end{equation}
where $\nabla^2 K_3 =0$ on $\IR^3$.   The vector potential, $\beta$, is then given by:
\begin{equation}
\beta ~\equiv~ \frac{K_3}{V}  \, (d\psi ~+~ A) ~+~  \vec{\xi} \cdot  d \vec y \,,
\label{Bpot}
\end{equation}
where
\begin{equation}
\vec  \nabla \times \vec \xi  ~=~ - \vec \nabla K_3 \,.
\label{xidefn}
\end{equation}

The one-form
\begin{equation}
\alpha ~\equiv~ V \, (dv + \beta) ~=~   V\, (dv +  \xi) ~+~ K_3 (d\psi + A)   \,,
\label{alphadefn}
\end{equation}
appears throughout the metric and flux and will play a significant role in that it has a manifest symmetry between the $v$ and $\psi$ fibers and the flux potential and the GH potential.  This symmetry lies at the heart of the spectral flow transformations \cite{Bena:2008wt}.  In particular, note that $\alpha \to -\alpha$  under the mapping
\begin{equation}
  V ~\leftrightarrow~ K_3\,;  \qquad  v \to  - \psi \,, \ \  \psi  \to  - v  \,; \qquad   A \to  - \xi \,, \ \  \xi  \to- A \,,
\label{SpecInv}
\end{equation}
The negative signs are required so as to respect the relations (\ref{AVreln}) and (\ref{xidefn}) between the harmonic functions and the vector potentials.  We will refer to the transformation, (\ref{SpecInv}), as { \it spectral inversion}.    

If one rewrites the solution by making the interchange above, one must also send $u \to -u$ so as to preserve the terms of the form $\alpha du$ in the metric and the electric potential terms in the flux.  One can then rewrite the entire supersymmetric form in terms of functions $\widetilde Z_I$, $\tilde \mu$ and a one form  $\tilde \omega$.   A straightforward calculation akin to that of \cite{Bena:2008wt} shows that 
\begin{align}
 \widetilde Z_i  & ~=~ \frac{V}{K_3} \, Z_i \,, \ \ i =1,2 \,;  \qquad \qquad  \widetilde Z_3 ~=~ \frac{K_3}{V}\,Z_3  ~+~  \frac{Z_1 Z_2}{K_3}  - 2\, \mu   \,, \nonumber \\
 \tilde k &~=~  - k  \,,  \qquad  \tilde \mu   ~=~  - \frac{V}{K_3} \, \mu ~+~   \frac{Z_1 Z_2 V }{K_3^2} \,;  \qquad \tilde \omega ~=~ - \omega \,.
\label{FunInv}
\end{align}
Under this transformation, the magnetic fluxes, $\Theta_j$  are also mapped to $\widetilde\Theta_j  = - \Theta_j$ because of the flip in the sign of $\alpha$. 
  
The goal is to use this  symmetry between two fibers to generate new classes of solutions and to formulate the theory in such a manner as to make  this symmetry more apparent  in BPS conditions.  

\subsection{The simplified BPS equations}

With our choice of background geometry, the BPS equations simplify significantly.  The equations for the fluxes and potentials reduce to
\begin{equation}
D \Theta_2 ~=~   *_4  D  \dot Z_1 \,, \qquad D *_4  D Z_1 = -  \Theta_2 \wedge  \Theta_3 \,,   \label{Theta2Z1} 
\end{equation}
and
\begin{equation}
D  \Theta_1~=~   *_4  D \dot Z_2  \,, \qquad  D *_4  D Z_2   = -  \Theta_1 \wedge  \Theta_3  \,.    \label{Theta1Z2}
\end{equation}
The equations for $k $ and $Z_3$ become:
\begin{eqnarray}
*_4 D *_4 (D Z_3 + 2\, \dot{k }  )   &=&       
 - *_4   \Theta_1 \wedge \Theta_2 ~+~ 2\, \big[ Z_1 \partial_v^2 Z_2 +  Z_2 \partial_v^2 Z_1 +  (\partial_v Z_1)(\partial_v Z_2) \big] \,,  \label{Z3simp} \\
D k  ~+~  *_4 D k     &=&  \sum_{I=1}^3 \, Z_I\,    \Theta_I     \label{simpangmomeqn} \,.
\end{eqnarray}
%

\subsection{The five-dimensional solutions}

\subsubsection{The ``classic'' solutions}

To motivate the construction of the six-dimensional solutions, it is useful to begin by recalling the well-known form of the five-dimensional BPS solutions.  These solutions are independent of the $GH$ fiber coordinate, $\psi$ \cite{Gauntlett:2004qy, Bena:2005va,Berglund:2005vb, Bena:2005ni, Saxena:2005uk}. 

The fluxes, $\Theta_j$ are harmonic and are given by expressions of the form (\ref{harmtwoform}):
\begin{equation}
\Theta_J~ =~ - \sum_{a=1}^3 \, \big(\partial_a \big( V^{-1}\, K_J \big)\big)  \, \Omega_+^{(a)}  \,,  \qquad J=1,2,3  \,,
\label{Thetaj}
\end{equation}
where $\nabla^2 K_j =0$ on $\IR^3$.    

The potentials, $Z_I$, are given by 
\begin{equation}
Z_I ~ =~  \frac{K_J \, K_K}{V} ~+~ L_I  \,,  \qquad I=1,2,3  \,,
\label{ZIfiveD}
\end{equation}
where $\{I, J, K\} = \{1, 2, 3\}$ are all distinct and where $\nabla^2 L_I =0$ on $\IR^3$.  The angular momentum vector has the form:
\begin{equation}
k ~ =~  \mu (d \psi + A) ~+~ \vec \omega \cdot d \vec y\,,
\label{kform}
\end{equation}
with
\begin{equation}
\mu ~ =~    \frac{K_1 \, K_2 \, K_3}{V^2} ~+~ \frac{1}{2} \, \sum_{I=1}^3 \,  \frac{K_I \, L_I}{V} ~+~M    \,,
\label{mufiveD}
\end{equation}
and $\nabla^2 M =0$ on $\IR^3$.  The angular momentum on $\IR^3$ is then given by
\begin{equation}
 \vec{\nabla} \times \vec{\omega}  ~=~ V\vec{\nabla} M ~-~ M\vec{\nabla} V  +  \frac{1}{2}\, \sum_{I=1}^3 \,  (K_I\vec{\nabla} L_I ~-~ L_I\vec{\nabla} K^I)\,,  
\label{omfiveD}
\end{equation}
and (\ref{mufiveD}) guarantees the integrability of this equation for $\vec{\omega}$.

The fact that these solutions are both $\psi$ and $v$ independent means that their form must be invariant under spectral inversion (\ref{SpecInv}).  Indeed, the transformation (\ref{FunInv}) can be rewritten as:
\begin{align}
 \widetilde V  ~=~  & K_3 \,, \quad   \widetilde K_3  ~=~  V \,,   \quad \widetilde K_1  ~=~ L_2\,, \quad  \widetilde K_2  ~=~ L_1 \,,   \nonumber \\
  \widetilde L_1  ~=~ & K_2\,, \quad  \widetilde L_2  ~=~ K_1 \,,   \quad \widetilde L_3  ~=~ - 2\, M \,, \quad  \widetilde M   ~=~  -\coeff{1}{2}\, L_3 \,. \label{KLMint}
 \end{align}
%

\subsubsection{More general five-dimensional solutions}
\label{Sect:MoreSols}

The solutions above are independent of both $v$ and $\psi$ and since the most general, five-dimensional BPS solutions have to satisfy (\ref{Theta5d})--(\ref{BPSfiveD4}), it is thus natural to ask about generalizations that are still independent of $v$ but depend upon $\psi$. 

If the base metric, $ds^2_4$, is smooth and Euclidean, it is easy to see that (\ref{harmtwoform}) represents the most general smooth solution to (\ref{Theta5d}).  Specifically, the equations,  (\ref{Theta5d}), imply that  the $\Theta_j$ are harmonic and the possible choices for $K_J$ given by: 
\begin{equation}
K_J  ~=~ k_{J,0} ~+~ \sum_{i=1}^N \, \frac{k_{J,i}}{|\vec y - \vec y^{(i)}| } \,, 
\label{Kform}
\end{equation}
for some parameters, $k_{J,i}$, form a basis for the harmonic forms.  Thus, in five-dimensions, the $\Theta_J$ are necessarily $\psi$-independent.  

One can see this more explicitly by taking $\Theta_j$ and subtracting its harmonic part,  $\Theta_{j, harm}$ to yield $\Theta'_j \equiv \Theta_j -\Theta_{j, harm}$. This is necessarily exact and so $\Theta'_j \wedge \Theta'_j$ (no sum on $j$) is all exact.    Hence
\begin{equation}
  0 ~=~  \int_\cB  \Theta'_j \wedge \Theta'_j ~=~ \int_\cB \, \Theta'_j \wedge *_4\Theta'_j   \,,
\end{equation}
However, the last integrand is necessarily non-negative and so one must have  $\Theta'_j \equiv 0$.  

There is, however, a gap in this argument for general BPS backgrounds:  If the base space is ambipolar \cite{Giusto:2004kj, Bena:2005va,Berglund:2005vb, Bena:2007kg}.  then the metric, $ds_4^2$, is singular and so  the Hodge decomposition theorem no longer applies. It is therefore quite possible that in ambipolar bases the Maxwell fields, $\Theta_j$, might be able to have a $\psi$-dependence.  This dependence would, however, have to be sourced in some manner associated with the critical surfaces where $V$ vanishes.  We will, however, not pursue this possibility here.

The simplest way to generate five-dimensional solutions that depend upon $\psi$ was discussed in \cite{Bena:2010gg}.  These solutions are important because they represent  an infinite family of smooth microstate geometries in six dimensions and thus can be used to generate  smooth microstate geometries in five dimensions by spectral flow \cite{Bena:2008wt}.   These solutions start with the fluxes exactly as in (\ref{Thetaj}) and (\ref{Kform}) and then introduce the $\psi$-dependence in the next layer of BPS equations by letting the functions, $L_I$, in (\ref{ZIfiveD}) depend upon $\psi$.   Equations (\ref{BPSfiveD1})--(\ref{BPSfiveD3}) then imply that the $L_I$ must be harmonic in four dimensions:
\begin{equation}
\nabla^2_{(4)} L_I ~=~ 0 \,. \label{harmLI}
\end{equation}

The last BPS equation, (\ref{BPSfiveD4}), can now be written as 
\begin{equation}
( \mu \vec {\mathcal{D}} V - V\vec {\mathcal{D}}  \mu  ) ~+~      \vec {\mathcal{D}}  \times \vec \omega ~+~    V \partial_\psi  \vec \omega  ~=~ -  V\, \sum_{I=1}^3 \, Z_I \, \vec \nabla \big(V^{-1} K^I  \big) \,,
\label{simpcovkeqn}
\end{equation}
where 
\begin{equation}
\vec {\mathcal{D}} ~\equiv~ \vec \nabla ~-~   \vec A \,\partial_\psi   \,. 
\label{simpcovD}
\end{equation}
The BPS equation, (\ref{BPSfiveD4}), has a gauge invariance:  $k \to k + df$ and this reduces to:
\begin{equation}
\mu \to \mu ~+~ \partial_\psi f \,, \qquad  \vec \omega \to  \vec \omega ~+~ \vec {\mathcal{D}} f \,. 
\label{fgaugetrf}
\end{equation}
It is simplest to use a Lorentz gauge-fixing condition, $d\star_4k =0$, which reduces to
\begin{equation}
V^2 \, \partial_\psi \mu   ~+~\vec {\mathcal{D}}  \cdot \vec \omega  ~=~ 0 \,.
\label{Lorgauge}
\end{equation}

The four-dimensional Laplacian can be written:
\begin{equation}
\nabla^2_{(4)}  F ~=~ V^{-1} \big[ V^2 \, \partial_\psi^2  F   ~+~ \vec {\mathcal{D}}  \cdot  \vec {\mathcal{D}}    F \big] \,.
\label{Lapl}
\end{equation}
Now take the covariant divergence, using $\vec {\mathcal{D}}$,  of (\ref{simpcovkeqn}) and use the Lorentz gauge choice, and one obtains:
\begin{equation}
V^{^2}\, \nabla^2_{(4)}    \mu  ~=~   \vec {\mathcal{D}} \cdot \Big( V\, \sum_{I=1}^3 \, Z_I \, \vec {\mathcal{D}} \big(V^{-1} K^I  \big) \Big) \,.
\label{mueqn}
\end{equation}
Remarkably enough, this equation is still solved by:
\begin{equation}
\mu ~=~  V^{-2} K_1 K_2 K_3~+~ \coeff{1}{2}\, \sum_{I=1}^3 V^{-1} K_{I}L_{I}  ~+~  M\,,
\label{muform}
\end{equation}
where, once again, $M$ is a harmonic function in four dimensions. Finally, we can use this solution back in (\ref{simpcovkeqn}) to simplify the right-hand side to obtain:
\begin{equation}
\vec {\mathcal{D}} \times \vec \omega ~+~ V \partial_\psi \vec \omega ~=~ V \vec {\mathcal{D}} M - M\vec {\mathcal{D}} V +\frac{1}{2} \, \sum_{I=1}^3 \big( K^{I} \vec {\mathcal{D}} L_{I} - L_{I}   \vec {\mathcal{D}} K^{I} \big).
\label{simpomegaeqn}
\end{equation}
Once again one sees the emergence of the familiar symplectic form on the right-hand side of this equation.  One can also verify that the covariant divergence (using $\vec {\mathcal{D}}$) generates an identity that is trivially satisfied as a consequence of  (\ref{AVreln}),  (\ref{Lorgauge}),  (\ref{muform})  and
\begin{equation}
 \nabla^2_{(4)}  L_I ~=~  \nabla^2_{(4)}    M ~=~ 0 \,.
\label{harmonicLM}
\end{equation}
%

\subsubsection{Spectral inversion revisited}
\label{Sect:SpecInv}

One can now use the spectral interchange symmetry generated by (\ref{SpecInv}) and realized in (\ref{FunInv}) to convert the $\psi$-dependent solutions into new $v$-dependent solutions of the BPS equations.  In particular, the generalization of (\ref{FunInv}):
\begin{equation}
K_3 \,  \widetilde Z_1   ~=~ V  \, Z_1  ~=~  K_3  \, K_2 ~+~ V  \, L_1 \,,  \qquad   K_3 \,  \widetilde Z_2   ~=~ V  \, Z_2  ~=~  K_3  \, K_1 ~+~ V  \, L_2 \,,
\label{Zrelns}
\end{equation}
now means that the new solutions have $v$-dependent fluxes governed by:
\begin{equation}
\widetilde  K_1 (v, \vec y)     ~=~    \, L_2 (\psi  , \vec y)\big|_{\psi = -v}    \,,  \qquad  \widetilde  K_2 (v, \vec y)     ~=~    \, L_1(\psi  , \vec y)\big|_{\psi = -v}    \,.
\label{KfromL}
\end{equation}
These solutions will then obey the more general BPS equations (\ref{Theta2Z1})--(\ref{simpangmomeqn}).  It is also important to note that because  (\ref{SpecInv}) is simply induced by a coordinate change, the new solution will also be smooth.  

One can now hybridize this observation with the strategy of Section \ref{Sect:MoreSols}.  That is, one can start by using the new $v$-dependent fluxes and then, once again, 
allow the solutions to develop a $\psi$-dependence, allowing the $L$ functions to depend upon $\psi$ as well as $\vec y$.  One then solves the remaining BPS equations.  Rather than pursue this course here we use this to motivate a significantly more general class of solutions that will be developed in the nest section.

\section{Families of doubly-fluctuating solutions}
\label{Sect:DoubleFluct}

Based on the observations in the previous section, it is relatively easy to formulate an Ansatz that will capture at least all the solutions proposed in Section \ref{Sect:SpecInv}.  We will see that it captures a far more general class of solutions.  

We will keep the geometry exactly as in Section \ref{Sect:GHbase}: A GH base with the six-dimensional geometry being completely independent of $v$. The background therefore has the spectral inversion symmetry (\ref{SpecInv}).  We also introduce  a generalization of the operator (\ref{simpcovD}): 
\begin{equation}
\vec {\mathcal{D}} ~\equiv~ \vec \nabla ~-~   \vec A \,\partial_\psi ~-~   \vec \xi \,\partial_v   \,,
\label{gencovD}
\end{equation}
and  define the second order operator:
\begin{equation}
{\cal L}  F ~\equiv~ \vec {\mathcal{D}} \cdot \vec {\mathcal{D}} \, F   ~+~ (V \, \partial_\psi - K_3 \, \partial_v)^2 F  \,. 
\label{cLdefn}
\end{equation}
Both of these operators are invariant under (\ref{SpecInv}).   Also note that 
\begin{equation}
{\cal L}  F  ~=~ - V\, *_4 D *_4 D\, F  \,,
\label{cLreln}
\end{equation}
where $D$ is the operator defined in (\ref{Ddefn}).

The appearance of the operator (\ref{cLdefn}) is very easy to understand because it is essentially the six-dimensional Laplacian for the metric  (\ref{sixmet}) acting on $u$-independent functions.  That is, one can easily verify that
\begin{equation}
\nabla_{(6)}^2 F(v,\psi, \vec y) ~=~  - \frac{1}{H \, V}  \, {\cal L}  F(v,\psi, \vec y)   \,.
\label{sixDLap}
\end{equation}
%

\subsection{The first layers of BPS equations}

We generalize the expression for fluxes, (\ref{Thetaj}), to
\begin{equation}
\Theta_j~ =~ - \sum_{a=1}^3 \, \big({\mathcal{D}}_a \big( V^{-1}\, K_j \big)\big)  \, \Omega_+^{(a)}  \,,  \qquad  j=1,2  \,,
\label{newThetaj}
\end{equation}
where, in principle, $K_j$ is a function of $v, \psi$ and $\vec y$. We also use, without loss of generality, the Ansatz
\begin{equation}
Z_1~ =~  \frac{K_3 \, K_2}{V} ~+~ L_1 \,,  \qquad Z_2 ~ =~  \frac{K_3 \, K_1}{V} ~+~ L_2 \,,  
\label{newZI}
\end{equation}
where the functions, $L_j$, are, as yet, general functions of $v, \psi$ and $\vec y$.   For completeness, the appendix to this paper contains the analysis of the BPS equations for the most general form of the $\Theta_j$.

If one substitutes this Ansatz into  the first BPS equations (\ref{Theta2Z1}) and (\ref{Theta1Z2}) one obtains the following linear equations for  $K_j$ and $L_j$, $j=1,2$: 
\begin{align}
&{\cal L} K_j ~=~  {\cal L} L_j ~=~   0 \,,  \qquad j=1,2 \,,   \label{KLeqns} \\ 
 & \partial_\psi K_1  ~+~ \partial_v L_2 ~=~   0 \,,  \qquad  \partial_\psi K_2 ~+~  \partial_v L_1 ~=~   0 \,,  \label{KLconstraint} 
\end{align}
where ${\cal L}$ is defined in (\ref{cLdefn}).

The general solution to the constraints  (\ref{KLconstraint}) is simply:
\begin{align}
& K_j ~=~  \hat K_j (v, \vec y)~+~ \partial_v H_j  (v, \psi, \vec y) \,,  \qquad j=1,2 \,,   \label{KForms} \\ 
& L_1 ~=~  \hat L_1 (\psi, \vec y)~-~ \partial_\psi H_2  (v, \psi, \vec y) \,,  \qquad   L_2 ~=~  \hat L_2 (\psi, \vec y)~-~ \partial_\psi H_1  (v, \psi, \vec y) \,,   \label{LForms} 
\end{align}
where
\begin{equation}
{\cal L} \hat K_j ~=~  {\cal L} \hat L_j ~=~  {\cal L}  H_j ~=~   0     \,. 
\label{KLHeqns}
\end{equation}
Note that the parts of the solution that involve $\hat K_j$ and $\hat L_j$ are actually redundant because they can be absorbed into various zero-mode parts of the $H_j$. We  have specifically exhibited $\hat K_j$ and $\hat L_j$ here because they represent precisely what would have been generated by the procedure outlined in Section \ref{Sect:SpecInv}.  This also makes it evident that the functions, $H_j$, represent something completely new and much more general.  We will discuss this more below.

\subsection{The last layer of BPS equations}

Without loss of generality, we can, once again,  use the Ans\"atze:
\begin{equation}
Z_3~ =~  \frac{K_1 \, K_2}{V} ~+~ L_3 \,,  \qquad \qquad   k ~ =~  \mu (d \psi + A) ~+~ \vec \omega \cdot d \vec y\,,
\label{kZ3form}
\end{equation}
with
\begin{equation}
\mu ~ =~    \frac{K_1 \, K_2 \, K_3}{V^2} ~+~ \frac{1}{2} \, \sum_{I=1}^3 \,  \frac{K_I \, L_I}{V} ~+~M    \,,
\label{muagain}
\end{equation}
where $L_3$ and $M$ are general functions that depend upon $(v, \psi, \vec y)$. 

Equation (\ref{simpangmomeqn}) for the angular momentum vector reduces to a straightforward generalization of  (\ref{simpomegaeqn}):
\begin{equation}
\vec {\mathcal{D}} \times \vec \omega ~+~ (V \partial_\psi - K_3  \partial_v) \vec \omega ~=~ V \vec {\mathcal{D}} M - M\vec {\mathcal{D}} V +\frac{1}{2} \, \sum_{I=1}^3 \big( K^{I} \vec {\mathcal{D}} L_{I} - L_{I}   \vec {\mathcal{D}} K^{I} \big).
\label{genomegaeqn}
\end{equation}

To simplify this and  (\ref{Z3simp}) we need to make a suitable generalization of the gauge choice (\ref{Lorgauge}).  To express this gauge, it is useful to introduce the spectral inversion of the function, $\mu$, under the effect of (\ref{KLMint}):
\begin{equation}
\tilde \mu ~\equiv~    \frac{L_1 \, L_2 \, V}{K_3^2} ~+~ \frac{1}{2} \, \left( \frac{K_1 \, L_1 +K_2 \, L_2  }{K_3} \right) \,  ~-~ \frac{V \, M}{K_3} ~-~\frac{1}{2} \,  L_3    \,,
\label{tildemu}
\end{equation}
and define
\begin{align}
\Phi ~\equiv~&  V^2 \, \partial_\psi \mu  + K_3^2 \, \partial_v  \tilde \mu  \nonumber \\
~=~&   K_3 \partial_\psi (K_1 K_2)  ~+~ V  \partial_v (L_1 L_2)  ~+~ \coeff{1}{2}\,(V  \partial_\psi + K_3 \partial_v) (K_1 L_1 +K_2 L_2)   \nonumber \\ 
& +  (V  \partial_\psi - K_3 \partial_v) (\coeff{1}{2}\,  K_3 L_3 + V M) \,.
\label{Phidef}
\end{align}
The gauge choice that simplifies all the equations is to take 
\begin{equation}
\vec {\mathcal{D}}  \cdot \vec \omega ~+~ \Phi ~=~ 0   \,,
\label{goodgauge}
\end{equation}
but we will not impose this and we will retain $\Phi$ in our equations.

If one uses the equations (\ref{KLeqns}) and  (\ref{KLconstraint})  one finds that   (\ref{Z3simp}) collapses to
\begin{equation}
{\cal L} L_3 ~=~ - 2\, \partial_v \big[ \vec {\mathcal{D}}  \cdot \vec \omega ~+~ \Phi  \big]    \,,
\label{L3eqn}
\end{equation}
while the covariant divergence of (\ref{genomegaeqn}) becomes:
\begin{equation}
\coeff{1}{2}\, K_3 \, {\cal L} L_3  ~+~ V \, {\cal L} M~=~ (V  \partial_\psi - K_3 \partial_v)   \big[ \vec {\mathcal{D}}  \cdot \vec \omega ~+~ \Phi  \big]    \,.
\label{genIntCond}
\end{equation}
Combining this with  (\ref{L3eqn}) yields
\begin{equation}
{\cal L} M~=~  \partial_\psi \big[ \vec {\mathcal{D}}  \cdot \vec \omega ~+~ \Phi  \big]    \,.
\label{Meqn}
\end{equation}

Therefore, with the gauge choice (\ref{goodgauge}), the functions $K_j$ and $L_j$ are fixed by  (\ref{KForms}),  (\ref{LForms}) and (\ref{KLHeqns}) and the remaining parts of the solution are given by (\ref{kZ3form}), (\ref{muagain}) and (\ref{genomegaeqn}) where $L_3$ and $M$ are general functions of $(v,\psi,\vec y)$ satisfying
\begin{equation}
{\cal L} L_3 ~=~  {\cal L} M ~=~ 0   \,.
\label{LMharm}
\end{equation}
These conditions then guarantee the integrability of (\ref{genomegaeqn}) for $\vec \omega$.

\subsection{The metric and its regularity}
\label{Sect:metreg}

One can write the metric in a  more symmetric form that is manifestly invariant under spectral inversion.  First recall that  in (\ref{alphadefn}) we defined the one form: 
\begin{equation}
\alpha ~\equiv~ V \, (dv + \beta) ~=~   V\, (dv +  \xi) ~+~ K_3 (d\psi + A)   \,.
\end{equation}
Now define  the functions: 
\begin{equation}
\hat H   ~\equiv~  \sqrt{(V Z_1)(V Z_2)}   \,, \qquad \cQ ~\equiv~   Z_1 Z_2 Z_3 V ~-~ \mu^2 V^2\,, \label{HQdefns}
\end{equation}
and the one form
\begin{equation}
\gamma ~\equiv~  K_3^2 \, \tilde \mu\,  (d\psi + A)  ~-~   V^2 \, \mu \, (dv +  \xi)  \,.\label{gammadefn}
\end{equation}
All of these quantities are invariant under the spectral inversion  transformation (\ref{KLMint}) and the quantity, $\cQ$, is simply  the $E_{7(7)}$ quartic invariant constructed out of the functions $V, K_I, L_I$ and $M$ \cite{Bena:2007kg}: 
\begin{align}
\cQ ~=~&  - V^2\,M^2   - 2 \, {K_1}\,{K_2}\,{K_3}\,M - M\,V\,\sum_{I=1}^3 \, {K_I}\,{L_I}
- \frac{1}{4} \,\sum_{I=1}^3 \, (K_I L_I)^2 \nonumber \\
&  \quad + V \, L_1\, L_2\, L_3 + (K_1 K_2 L_1 L_2+K_1 K_3 L_1 L_3+K_2 K_3 L_2 L_3) \,. 
\label{QasEseven}
\end{align}
The six dimensional metric, (\ref{sixmet}) can now be written as:
\begin{equation}
ds_6^2 ~=~   2 \hat H^{-1} \, (du +  \omega )\, \alpha ~-~ \hat H^{-3} \,\big[ \cQ\, \alpha^2 ~+~ \gamma^2  \big] ~-~   \hat H \, d \vec y \cdot d \vec y \,.   \label{newsixmet}
\end{equation}

The standard, bubbled microstate geometries   \cite{Bena:2005va,Berglund:2005vb,Bena:2007kg,Bena:2008wt,Bena:2008dw,Saxena:2005uk}  allow singularities at points in the  $\IR^3$ defined by $\vec y$.  Indeed, near such a singular point, $P$,  one has $V  ~\sim~ \frac{ q_p}{r_p}$ while one also requires that the $Z_I$ are finite as $r_p \to 0$ and the bubble equations require that $\mu(r_p) =0$.  This means that, as $r_p \to 0$, one has
\begin{equation}
\hat H\,, \cQ\,, \alpha \,,  \gamma ~\sim~  \cO\big(  r_p^{-1}  \big) \,.  \label{singstrength}
\end{equation}
Since supertubes can be mapped onto microstate geometries by spectral flow \cite{Bena:2008wt}, it follows that supertube  solutions have identical asymptotics to that of (\ref{singstrength}).  One can also check this directly.
It then follows that in all such configurations the  metric (\ref{newsixmet}) remains smooth (up to orbifold points).  The apparent singularity on the $\IR^3$ base can be resolved by the standard coordinate change: $r_p = \frac{1}{4} R^2$.  

For supertubes one can also get smooth solutions with slightly more singular behavior than that  allowed by (\ref{singstrength}) \footnote{We are very grateful to Stefano Giusto for pointing this out and helping to clarify this point.}.  Indeed, $\cQ$ and $\gamma$  can have double poles that cancel in (\ref{newsixmet}) so that the metric remains regular.  The simplest, and perhaps only, examples of this are coordinate transformations of solutions that satisfy (\ref{singstrength}) but in which constants in $\cF = -Z_3$ are  removed via a re-definition of $u$.  This can, in turn, move singular terms between the functions that define the solution. To fix this ambiguity we will require that our  asymptotically flat solutions have $Z_3 \to c_3$ at infinity, where $c_3 \ne 0$ is a constant.  Upon completing the square in (\ref{sixmet}) one can then rewrite the six-dimensional metric as: 
\begin{equation}
ds^2_6  ~=~ -Z_3  H^{-1}  \big[ dv+ \beta - Z_3^{-1} (du + k) \big]^{2} ~+~ (H Z_3)^{-1}  (du + k)^2  ~-~ H \, ds_4^2(\cB)\,.
\label{rediagsixmet}
\end{equation}
Since $Z_3$ and $H$ go to non-zero constants at infinity, this  means that in Kaluza-Klein reduction on the $v$ circle to five dimensions, $u$ becomes the time coordinate.  It is in this description that supertube solutions  are most easily related to five-dimensional bubbled solutions and both sets of solutions obey (\ref{singstrength}).   Henceforth, we will  assume that our solutions, including supertubes, obey (\ref{singstrength}).

Finally, it is important to note that near a singular point of the harmonic functions,  the metric on the $(v,\psi)$ torus has a pre-factor of  $\hat H^{-3} \sim r_p^3$ and, given the asymptotic behavior in  (\ref{singstrength}), this will pinch off the circle defined by $r_p \gamma$ while the circle defined by $r_p \alpha$ will remain finite. For the standard  five-dimensional bubbled microstate geometries this corresponds to pinching off the $\psi$-circle while the $v$-circle remains finite and for the standard supertube \cite{Bena:2008wt} the $v$-circle pinches off and the $\psi$-circle remains finite.

\subsection{The physical structure underlying the BPS system}

In conjecturing the existence of a superstratum \cite{Bena:2011uw}, one of the crucial first steps was to argue that this D1-D5-P system would retain the same supersymmetries  if the D1-D5 system were ``tilted and boosted''  so as to lay it out along an arbitrary closed profile in $(v,\psi,\vec y)$.  These configurations were then to be smeared along $v$ so as to make a supersheet and it was proposed that if this configuration also had a  KKM dipole charge that was arranged in the proper manner then the whole BPS configuration would remain smooth.    

In the simplest, standard D1-D5-P configuration, the common direction of the  D1-D5 system lies along the circle defined by $v$ while the momentum modes excite  oscillations in the transverse four-manifold described by $(\psi,\vec y)$.   If one tilts and boosts this configuration in the manner described in \cite{Bena:2011uw} then some of the D1 and D5 electric charges are tilted into $d1$ and $d5$ magnetic dipole charges and some of the momentum, $P$, is tilted into angular momentum around the profile.  

Now recall that the functions $Z_1$ and $Z_2$ encode the charge densities associated with the D1 and D5 branes respectively and so the pairs, $(L_1,K_2)$ and $(L_2,K_1)$, encode the $(electric, magnetic)$ charge densities of the D1 and D5 branes respectively.    The new feature of the solutions presented here are the functions, $H_j(v,\psi,\vec y)$,  appearing in (\ref{KForms}) and (\ref{LForms}).  These functions tie the D1-d1 and D5-d5 charges together in a manner that reflects precisely the tilting process described in \cite{Bena:2011uw} and the fact that this arises directly from the BPS conditions provides further support for the arguments given in \cite{Bena:2011uw}.    It should, of course, be remembered that we have frozen the background geometry so that it is independent of $(v, \psi)$ and so the shape of the profile that we are trying to generate does not fluctuate directly.  Instead, we are fluctuating the charge densities within the $(v, \psi)$-independent profile and it is these densities that are being tilted and boosted. The effect of these fluctuating densities back-react in the full metric and will thus change the physical size of the configuration as a function of $(v, \psi)$ and so the shape will indeed ultimately fluctuate.

The fact that the functions, $H_j(v,\psi,\vec y)$, are general solutions of the reduced Laplacian (\ref{sixDLap}) is a very natural generalization of the harmonic charge sources  that are part of the five-dimensional system.  Indeed, one should recall that if there is a supersymmetry, $\epsilon$, then one can construct the vector\footnote{The supersymmetry may have internal indices and this expression may involve some contractions over these indices.}:
\begin{equation}
T^\mu~\equiv~     \bar \epsilon \gamma^\mu \epsilon 
\label{Sandwich}
\end{equation}
and this will generically be a time-like, or null Killing vector.  In five dimensions this is time-like and hence the BPS solutions have a time-translation invariance.  However, in six dimensions, $T^\mu$ is a null vector \cite{Gutowski:2003rg} and this accounts for the $u$-independence of the solution.  The fact that $T^\mu$ is null means that the hypersurfaces of constant $u$ are null and the induced metric on these surfaces is degenerate.  

This accounts for the somewhat degenerate form of (\ref{cLdefn}): While it involves derivatives with respect to small five spatial variables,  $(v,\psi,\vec y)$, it is written in a diagonal form that is only the sum of four squares.     This fact will have a significant impact on the space of solutions.  Indeed, one can imagine trying to find the Green functions for ${\cal L}$ by following the approach of \cite{Bena:2010gg} and integrating out the time direction in the propagator of the full Laplacian.  In six-dimensions, this will involve integrating out the null coordinate $u$ in the Green function for the six-dimensional Laplacian.  This is not a well-defined procedure in the six-dimensional theory because the null initial-value problem  may not be well-posed and one will potentially be integrating along singularities corresponding to propagating data.

\section{Multi-centered configurations}
\label{Sect:Multi}

Multi-centered are those solutions that start by taking a multi-centered geometry in which $V$ and $K_3$ have the form (\ref{AVreln}) and (\ref{Kform}).  Initially, we will not make any assumptions about the form of the other functions, $K_j, L_j$, $(j=1,2)$, $L_3$ and $M$. We start by  analyzing, in detail, one of the simplest, non-trivial microstate geometries:  the two-centered solution and then use this  to describe what we believe will be the structure of a generic multi-centered solution. 

\subsection{The general two-centered configuration and  $AdS_3 \times S^3$}
\label{Sect:AdStimesS}

To define the general two-centered solution it is simplest to introduce  cylindrical polar coordinates, $(z, \rho, \phi)$,  on the $\IR^3$ base and define 
\begin{equation}
r_\pm ~\equiv~   \sqrt{\rho^2 ~+~ (z\mp a)^2 } \,. \label{rpmdefn}
\end{equation}
The key geometric elements are then given by: 
\begin{eqnarray}
V &=&  \frac{q_+ }{r_+} ~+~  \frac{ q_- }{r_-}  \,, \qquad  A  ~=~ \Big(  q_+  \frac{(z-a)}{r_+} ~+~ q_-  \frac{ (z + a)}{r_-} \Big) \, d\phi    \,,    \label{KVtwoC}  \\
 K_3  &=&   \frac{ k_+ }{r_+} ~+~  \frac{ k_-}{r_-}    \,, \qquad  \xi ~=~    - \Big(k_+  \frac{ (z-a)}{r_+} ~+~  k_- \frac{ (z+a)}{r_-}  \Big) \, d\phi \label{AxitwoC}     \,,  
\end{eqnarray}
The two-centered system  is greatly simplified by working with bipolar coordinates:
\begin{equation}
\rho ~=~ a \, \sinh \xi \, \sin \theta \,, \qquad  z ~=~ a \, \cosh \xi \, \cos \theta \label{bipolar}     \,,
\end{equation}
and then one has:
\begin{equation}
r_\pm ~=~   a\, (\cosh \xi~\mp~ \cos \theta)    \,.
\end{equation}
One can then reparametrize the $T^2$ of $(v, \psi)$ by introducing the angles, $\chi$ and $\eta$, defined by:
\begin{align}
 \partial_\chi & ~=~   - (q_+  +  q_-) \partial_\psi + (k_+ +   k_-) \partial_v  \,, \qquad  \partial_\eta ~=~    (q_+ -  q_-) \partial_\psi - (k_+  -  k_-) \partial_v  \nonumber \\
 \Leftrightarrow \quad  \chi &~=~    \frac{1}{2\, \Delta} \, \big( (k_+  -  k_-)\,  \psi + (q_+ -  q_-)\, v \big)    \,,  \quad 
   \eta ~=~   \frac{1}{2\, \Delta} \, \big( (k_+ +   k_-)\,\psi  + (q_+  +  q_-) \,v \big)  \label{chietadefn}   \,,
\end{align}
where $\Delta \equiv (q_+ k_- - q_- k_+)$. 

The operator $\cL$ then has the relatively simple form:
\begin{align}
\cL F ~=~   a^{-2} (\sinh^2 \xi  + \sin^2 \theta)^{-2} \bigg[&  \frac{1}{\sinh \xi}\partial_\xi \big( \sinh \xi \, \partial_\xi  F\big) ~+~  \frac{1}{\sinh^2 \xi} (\partial_\eta +  \cosh \xi\, \partial_\phi)^2 F    \nonumber \\ 
&+  \frac{1}{\sin \theta}\partial_\theta \big( \sin \theta \, \partial_\theta  F\big)~+~ \frac{1}{\sin^2 \theta}  (\partial_\chi +  \cos \theta \, \partial_\phi)^2 F \bigg] \label{cLform}   \,.
\end{align}

The manifolds,  $S^3$ and $AdS_3$, can be thought of as a unit sphere in $\IC^2$ and as a unit hyperboloid in $\IC^{1,1}$  respectively:
\begin{align}
ds_{S^3}^2 & ~=~ |dw_1|^2 ~+~ |dw_2|^2 \,, \qquad   |w_1|^2 ~+~ |w_2|^2 ~=~ 1 \,; \\
ds_{AdS_3}^2 & ~=~ |dz_1|^2 ~-~ |dz_2|^2 \,, \ \ \qquad   |z_1|^2 ~-~ |z_2|^2 ~=~ 1 \,.
\end{align}
Parametrizing these surfaces in standard fashion:
\begin{align}
w_1 &~=~ \cos \coeff{1}{2} \theta  \, e^{i(\chi - \phi)}  \,, \quad w_2 ~=~ \sin \coeff{1}{2} \theta  \, e^{i(\chi + \phi)}  \,; \\
z_1 &~=~ \cosh \coeff{1}{2} \xi  \, e^{i(\eta - t)}  \,, \quad z_2 ~=~ \sinh \coeff{1}{2} \xi  \, e^{i(\eta + t)}   \,,
\end{align}
leads to the polar forms 
\begin{align}
ds_{S^3}^2 & ~=~    \coeff{1}{4} \, (d\theta^2 ~+~   \sin^2 \theta \, d\phi^2 ~+~ (d \chi -  \cos \theta\, d\phi)^2 ) \nonumber \\
& ~=~    \coeff{1}{4} \, d\theta^2 ~+~   \sin^2  \coeff{1}{2} \theta \, (d\chi+ d \phi)^2  ~+~   \cos^2 \coeff{1}{2} \theta \, (d\chi - d \phi)^2    \,; \label{S3met} \\
ds_{AdS_3}^2 & ~=~    \coeff{1}{4} \, (d\xi^2 ~+~   \sinh^2 \xi \, d\eta^2  ~-~  (d t -  \cosh \xi \, d\eta)^2 )  \nonumber  \\
&~=~    \coeff{1}{4} \, d\xi^2 ~+~   \sinh^2  \coeff{1}{2} \xi\, (d\eta+ d t)^2  ~-~   \cosh^2 \coeff{1}{2} \xi \, (d\eta - d t)^2  \,.  \label{AdS3met} 
\end{align}
The Laplacians on these spaces are $4 \cL_j$ where: 
\begin{align}
\cL_1 \, F  & ~\equiv~  \frac{1}{\sinh \xi}\partial_\xi \big( \sinh \xi \, \partial_\xi  F\big) ~+~  \frac{1}{\sinh^2 \xi}  (\partial_\eta +  \cosh \xi\, \partial_t)^2 F  ~-~ \partial_t^2  F   \,,  \label{AdS3Lap} \\
\cL_2 \, F  &~\equiv~    \frac{1}{\sin \theta}\partial_\theta \big( \sin \theta \, \partial_\theta  F\big)~+~ \frac{1}{\sin^2 \theta}  (\partial_\chi +  \cos \theta \, \partial_\phi)^2 F~+~   \, \partial_\phi^2    F     \,.\label{S3Lap} 
\end{align}

The important point is that the operator, $\cL$, that defines the solutions of interest is simply given by:
\begin{equation}
\cL ~=~  (\cL_1 + \cL_2)\big|_{\phi = t}   \,.
\end{equation}
Thus, even though the final six-dimensional metric is not ultimately going to be $AdS_3 \times S^3$, the differential operator, $\cL$, is precisely the Laplacian of $AdS_3 \times S^3$ acting on $u$-independent modes, where $u \equiv \phi - t$.

\subsection{Fluctuations}

\subsubsection{Generalities}

It is rather straightforward to see that there are no non-singular BPS fluctuations that fall off suitably rapidly at infinity on $AdS_3 \times S^3$.  Suppose that $F(\xi,\theta,\xi,\eta, \phi)$ were such a fluctuating mode and observe that:
\begin{align}
0 ~=~  - \int_{AdS_3 \times S^3} \,  &  a^2 (\sinh^2 \xi  + \sin^2 \theta)^2 \sinh^2 \xi \, (F \cL\ F)  \nonumber  \\
~=~  \int_{AdS_3 \times S^3} \,   &  \sin^2 \theta \big[   (\sinh \xi \, \partial_\xi  F)^2 ~+~   ( \partial_\eta F +  \cosh \xi\, \partial_\phi F )^2 \big] \nonumber \\ 
 & +~ \sinh^2 \xi   \big[ (\sin \theta \, \partial_\theta  F)^2~+~   ( \partial_\chi F +  \cos \theta \, \partial_\phi F )^2 )  \big]   \,.
\end{align}
where we have assumed that $F$ vanishes fast enough so that the boundary terms at infinity that behave as $\sim  e^{2 \xi}  F \partial_\xi F$ can be discarded.  All the terms in the integrand are positive definite and so they must all vanish.  This means that $F$ must be constant.   Given the fact that the general operator, $\cL$, in (\ref{cLdefn}) is the sum of squares, even when $V$ is ambipolar, one would expect a similar conclusion for a generic, multi-centered solution.

Thus fluctuating modes must either be non-normalizable or must have singularities.  While there might be interesting solutions that  involve the former, as we discussed in Section \ref{Sect:metreg}, earlier work shows there are huge families of smooth solutions in which the  harmonic functions are sourced on surfaces of spatial co-dimension $3$.  The simplest of these has singular surfaces that are points in $\IR^3$ and are swept out by $(v,\psi)$.    Thus the we will further specialize our notion of multi-centered solutions to those in which  the harmonic functions, $K_j,  L_j, L_3$ and $M$ have singularities of order $\cO(|\vec y - \vec y^{(i)}|^{-1} )$ at the points,  $\vec y^{(i)}$, where $V$ and $K_3$ are similarly singular.  However, unlike $V$ and $K_3$, the functions $K_j,  L_j, L_3$ and $M$ will be allowed to depend upon $(v, \psi)$.   As noted earlier,  the effect of the singular behavior of the harmonic functions involves  pinching off a direction in the $(v,\psi)$ torus, thereby creating a topological cycle that can then support non-trivial, smooth cohomological fluxes.  

We now examine perhaps the simplest example.

\subsubsection{A specific example}

We start with the solution in Section \ref{Sect:AdStimesS} with 
\begin{equation}
q_- ~=~ k_+ ~=~ 1 \,, \qquad  q_+ ~=~ k_- ~=~ 0 \label{simpchoice} \,.
\end{equation}
The base manifold, $\cB$,  is them simply flat $\IR^4 = \IC^2$ and the coordinate change relating this to the GH form is simply 
\begin{equation}
\zeta_1 ~=~  \sqrt{  r_-} \, \cos (\coeff{1}{2} \theta_-)  \, e^{i(\psi + \phi)/2} \,, \qquad  \zeta_2  ~=~    \sqrt{ r_-} \, \sin (\coeff{1}{2} \theta_-)  \, e^{i(\psi - \phi)/2} \,,
\end{equation}
where $(r_-, \theta_-, \phi)$ are polar coordinates with $r_-$  defined in  (\ref{rpmdefn}).  The Green function of the Laplacian $\IR^4$ for a source located at $r_+ =0$ is simply a constant multiple of $\Lambda$ where:
\begin{align}
\Lambda^{-1} ~=~  & \big | \zeta_1 -  \sqrt{2 a} \big|^2 ~+~ \big | \zeta_2 \big|^2  \\ 
~=~ &  (r_- + 2a)~-~2\, \sqrt{2 a r_-} \, \cos (\coeff{1}{2} \theta_-)  \, \cos \coeff{1}{2}( \psi + \phi)  \,.
\end{align}
One can then integrate this against any Fourier mode $e^{i m(\psi + \phi)/2}$ to get a fluctuating harmonic source.  This is an elementary contour integral and it yields:
\begin{equation}
F_m^+ ~=~   \frac{1}{r_+} \, \bigg( \frac{\cos (\coeff{1}{2} \theta)  }{\cosh (\coeff{1}{2} \xi)  }  \bigg)^{| m |}  \, e^{i m (\psi + \phi)/2} 
~=~  \frac{1}{a\, (\cosh \xi- \cos \theta)} \, \bigg( \frac{\cos (\coeff{1}{2} \theta)  }{\cosh (\coeff{1}{2} \xi)  }  \bigg)^{| m |}  \, e^{-i m (\chi + \eta- \phi)/2}  \label{Fmplus}  \,.
\end{equation}
where we have written the results in terms of the coordinates (\ref{bipolar}) and the angles $(\chi, \eta)$ defined in  (\ref{chietadefn}).  By construction, these are harmonic functions on the four-dimensional base and thus satisfy $\cL F_m^+ = 0$, for all values of $m$.

One can similarly verify that 
\begin{equation}
F_m^- ~\equiv~   \frac{1}{r_-} \, \bigg( \frac{\sin (\coeff{1}{2} \theta)  }{\cosh (\coeff{1}{2} \xi)  } \bigg)^{| m |}   \, e^{-i m (v  + \phi)/2}    
~=~  \frac{1}{a\,(\cosh \xi+ \cos \theta)} \, \bigg( \frac{\sin (\coeff{1}{2} \theta)  }{\cosh (\coeff{1}{2} \xi)  } \bigg)^{| m |}   \, e^{-i m (\chi - \eta + \phi)/2} \label{Fmminus}   
\end{equation}
also satisfy $\cL F_m^- = 0$.  However, these functions now have explicit mode dependence on $v$ and thus go beyond the standard harmonic Ans\"atze.  

The functions, $F_m^\pm = 0$, have several very important properties.  First, they are smooth except except a $\cO (r_\pm^{-1})$ singularity as  $r_\pm \to 0$.  The zero-modes are precisely $F_0^\pm = \frac{1}{r_\pm}$ and, for $m \ne 0$, one has:
\begin{equation}
F_m^+\big |_{\theta = \pi} ~=~ 0 \,, \qquad  F_m^-\big |_{\theta = 0} ~=~ 0  \,, \qquad m \ne 0 \label{niceprop} \,.
\end{equation}
Finally, the second expressions in (\ref{Fmplus}) and  (\ref{Fmminus})  show that the  $F_m^\pm$, are smooth as functions on $AdS_3 \times S^3$, independent of the choice  (\ref{simpchoice}).  One can therefore immediately generalize our discussion by working in $AdS_3 \times S^3$ and dropping the condition (\ref{simpchoice}) and leaving the parameters $q_\pm$ and $k_\pm$ completely generic.  It should be noted that we have not made a careful discussion of the proper periodicities of the angles, 
$\chi$ and $\eta$ and so one might be dealing with an orbifold of $AdS_3 \times S^3$.  We will, however, continue to impose (\ref{simpchoice}) so that we can easily relate our results to earlier work on bubbled geometries and supertubes.

One can now generate solutions by taking $H_j$, $L_3$ and $M$ to have Fourier expansions:
\begin{equation}
\sum_{m = -\infty}^\infty \, (b_m^+ \, F_m^+ ~+~ b_m^- \, F_m^-)    \,,
\end{equation}
where reality requires $b_{-m}^\pm = (b_m^\pm)^*$.  The fact that all the functions only have singularities of order $r_\pm^{-1}$ means that one can create completely smooth geometries.   At $r_+ =0$, the function $K_3$ is singular and the $v$ fiber and the $\phi$-circle pinch off.   This point corresponds to $\xi =0$, $\theta =0$ and so (\ref{niceprop}) implies that the only  non-trivial fluctuations come from  $F_m^+$ and so lie along the non-collapsing $\psi$ fiber.  The construction of regular solutions exactly follows the discussion of the wiggling supertubes in \cite{Bena:2010gg}.  As discussed in detail in  \cite{Bena:2010gg}, regularity near the supertube will impose constraints on the charge densities in the $H_j$, $L_3$ and $M$ and there will be one remaining, freely choosable charge density function, $\rho_+(\psi)$,  at this point.  Similarly, at $r_- =0$ or  to $\xi =0$,  $\theta =\pi$,  the function $V$ is singular and the $\psi$ fiber  and the $\phi$-circle  pinch off.   Again, (\ref{niceprop}) implies that the only  non-trivial fluctuations come from  $F_m^-$ and so lie along the non-collapsing $v$ fiber.  The regularity at $r_- =0$ will be the spectral inversion of the  regularity at $r_+ =0$ and one will be left with another freely choosable charge density function, $\rho_-(v)$.

To be more specific, one can find solutions with
\begin{equation}
K_j ~=~  \hat K_j (v, \vec y)  \,,  \ \   L_j ~=~  \hat L_j (\psi, \vec y)\,,   \quad j=1,2\,;   \qquad L_3  ~=~  \hat L_3 (v, \vec y) \,,    \qquad  M ~=~  M (\psi, \vec y) \,, 
\end{equation}
where a dependence on $(\psi, \vec y)$ implies an expansion in  $F_m^+$ alone and a dependence on $(v, \vec y)$ implies an expansion in  $F_m^-$ alone. In principle there are six freely choosable charge density  functions, three at each point: $\rho_J^+(\psi)$, $\rho_J^-(v)$.  At $r_+ =0$ one has $\theta =0$ and so the $K_j$ and $L_3$ collapse to their zero modes.  The analysis of regularity then exactly follows the analysis of   \cite{Bena:2010gg}, which means that the $\rho_J^+(\psi)$ can all be parametrized in terms of one function, $\rho^+(\psi)$.  At $r_- =0$ one has $\theta =\pi$ and so the $L_j$ and $M$ collapse to their zero modes and analysis of regularity is the spectral inversion of the analysis at  $r_+ =0$, which means that the $\rho_J^-(v)$ can all be parametrized in terms of one function, $\rho^-(v)$.  

If one removes the condition  (\ref{simpchoice}), and works with general $q_\pm, k_\pm$, then the foregoing discussion goes through as before except that the collapsing circles and density functions are parametrized by $(\phi, \chi \pm \eta)$ and $\chi \mp \eta$ respectively.  Using (\ref{chietadefn}), one obtains
\begin{equation}
\eta + \chi   ~=~      \frac{1}{\Delta} \, \big(  k_+  \,  \psi ~+~  q_+ \, v \big)    \,,  \qquad 
\eta  -  \chi~=~    \frac{1}{\Delta} \, \big(  k_- \,  \psi ~+~  q_-  \, v \big)    \label{pinchedcircles} \,,
\end{equation}
and these define the modes along the circles of finite size at $r_+ =0$ and $r_- =0$ respectively. This is, of course, consistent with the observation that  the finite circle and its modes are defined by $r_\pm \alpha$ as in Section \ref{Sect:metreg}.

\subsection{The general form of these solutions}
\label{Sect:gerform}

Perhaps the most important lesson of the last section is that in the multi-centered solutions one can have have special classes of singular sources in the solution and, at these sources, one circle in the $(v, \psi)$ torus pinches off and the source charge can then be spread in a  general line distribution along the other direction.  

This is also evident in the structure of the differential operator, $\cL$, defined in (\ref{cLdefn}).   The $(v, \psi)$ modes contribute to the following terms to this operator:
\begin{equation}
(V \, \partial_\psi - K_3 \, \partial_v) \,, \qquad (\vec A \,\partial_\psi ~+~   \vec \xi \,\partial_v)   \,,
\label{specterms}
\end{equation}
Suppose that  $V$ and $K_3$ have their generic forms (\ref{Vform}) and (\ref{Kform}) then the foregoing terms will have singularities of the form $\cO (|\vec y - \vec y^{(j)}|^{-1})$ except for modes $e^{i (n v + p\psi)}$ where the contributions for the two fibers cancel, that is, when: 
\begin{equation}
p\, q_j  ~-~ n\,  k_{3, j}  ~=~ 0\,.
\label{nicemodes}
\end{equation}
Thus the nature of the differential equation, and its solutions, will be quite different for generic modes and for special modes satisfying (\ref{nicemodes}). 
This identity implies $q_j   (n v + p\psi)  = n  (q_j   v + k_{3, j} \psi)$ and so  these special modes at $\vec y^{(j)}$ define the Fourier series of  functions of one variable that depend  upon $\sigma_j  \equiv (q_j   v + k_{3, j} \psi)$.  Now recall that the circle that remains of finite size is defined by $r_j \alpha \to q_j dv +  k_{3, j} \psi  = d\sigma_j$.

The modes satisfying this relationship lie along the circle that remains large at $\vec y^{(j)}$ and a linear charge distribution in these Fourier modes will only give rise to the required $\cO (|\vec y - \vec y^{(j)}|^{-1})$ singularity in the solutions to $\cL F =0$.  This identity implies $q_j   (n v + p\psi)  = n  (q_j   v + k_{3, j} \psi)$  and so this means that the modes at $\vec y^{(j)}$ must depend upon $\sigma_j  \equiv (q_j   v + k_{3, j} \psi)$.  Thus in a general solution we expect to be able to introduce  line sources at every point, $\vec y^{(j)}$, and the source densities will be functions of one variable, $\sigma_j$.  

We therefore expect that a generic fluctuating BPS solution based upon the Ansatz of Section \ref{Sect:GHbackground} can depend in a highly non-trivial manner on both variables $v$ and $\psi$, however this dependence is generated by source functions of one variable located at  the points $\vec y^{(j)}$.  The modes introduced at $\vec y^{(j)}$ depend upon the KKM and GH charges at that point and so by varying these charges between points one can get broad classes of fluctuations.  

\section{Conclusions}
\label{Sect:Conclusions}
 
We have analyzed the BPS equations of minimal six-dimensional supergravity coupled to one anti-self-dual tensor multiplet.  In particular, we have focussed upon a simple class of five-dimensional spatial backgrounds that may be thought of as $T^2$ fibration over a flat $\IR^3$ base.  This fibration is non-trivial because the fibration of the circles involves two independent sets of KKM's. The generic BPS configuration we considered could fluctuate with densities that depend freely on both directions of the torus, $T^2$.  However, we found that requiring smooth configurations restricts these densities to functions of one variable, albeit a different torus circle depending on each pair of  KKM charges.  Thus, by choosing different combinations of KKM charges one can obtain rich families of doubly fluctuating microstate geometries that depend  non-trivially on all directions within the $T^2$.    

It was conjectured that  general  superstratum  \cite{Bena:2011uw} will be a smooth solution of the supergravity theory studied here and yet has shape and density modes  that are general functions of two variables.  Our analysis here lends  support to the construction outlined in \cite{Bena:2011uw}.  In particular, the first step of this construction involves tilting and boosting the D1-D5-P system to generate d1-d5 dipole moments and angular momentum along a new profile.  We showed here that the six-dimensional BPS equations admit solutions that precisely represent this tilting and boosting procedure.   

The next and most difficult step in the construction of a generic superstratum is to add KKM's along the new profile so as to desingularize the tilted and boosted D1-D5-P system.  Here we have managed to realize this is a limited manner: 
Our solutions may be thought of as semi-rigid superstrata in that they are not sourced by generic functions of two variables. This seems to be a direct result of the rigidity of our array of Kaluza Klein monopoles: regularity at each KKM selects the  direction of the charge density dependence within the $T^2$.  If the KKM charge configuration could be made to vary non-trivially as a function of some combination of the $T^2$ fibers, $v$ and $\psi$, then the smooth configurations might indeed involve density functions that are generic functions of two variables.  This will, however, involve solving the  non-linear system (\ref{Theta3defn})  and (\ref{betacond}) for a general vector field, $\beta$.  While this is challenging, it may not be impossibly difficult because it is a form of self-dual Yang Mills equation that arises as the integrability condition of the linear system, (\ref{Jcond}),  for the $J^{(A)}$. 

Most of the focus of the latter part of this paper has been upon microstate geometries and smooth solutions.  One should not forget that there are very interesting singular solutions, like black holes and black rings.  The analysis of the BPS equations in the first sections of this paper will certainly provide interesting new families of such solutions in which there are fluctuations along the  $T^2$.  More generally, the ultimate simplicity of the BPS system based on the $T^2$ fibration suggests that it might be used as more general ``Floating-brane Ansatz'' as in \cite{Bena:2009fi}.  This might lead to six-dimensional generalizations of the whole class of almost-BPS and non-BPS configurations.

\bigskip
\leftline{\bf Acknowledgements}
\smallskip
We would like to thank  Iosif Bena, Stefano Giusto, Masaki Shigemori and Orestis Vasilakis for valuable discussions.   NPW is grateful to the IPhT, CEA-Saclay and to the Institut des Hautes Etudes Scientifiques (IHES), Bures-sur-Yvette, for hospitality while this work was completed.  This work was supported in part by the DOE grant DE-FG03-84ER-40168.  


\appendix
\section{Fully General Equations}

\renewcommand{\theequation}{A.\arabic{equation}}
\renewcommand{\thetable}{A.\arabic{table}}
\setcounter{equation}{0}
\label{appendixA}

In Section \ref{Sect:DoubleFluct}, we made certain simplifying assumptions that result in a system of equations constrained by \eqref{KLconstraint}.  This gives a particularly simple set of equations $\mathcal{L} f = 0$, where $f$ is any of $K_1, K_2, L_1, L_2, L_3, M$.  However, this is not the most general form of the equations.  Revisiting \eqref{newThetaj}, we can easily make a  general Ansatz at least for the $\Theta_j$ and do it in a manner that leads to a similar simplification of the source terms.  We simply introduce  vector fields, $\vec \lambda_{j}$, into (\ref{newThetaj}):
\begin{equation}
\Theta_j~ =~ - \sum_{a=1}^3 \, \Big({\mathcal{D}}_a \big( V^{-1}\, K_j \big) + \lambda_{j \, a} \Big)  \, \Omega_+^{(a)}  \,,  \qquad  j=1,2  \,.
\end{equation}
Then the constraints \eqref{KLconstraint} are replaced by
\begin{equation}
\begin{split}
\vec{\mathcal{D}} ( \partial_\psi K_1 + \partial_v L_2 ) &= \vec{\mathcal{D}} \times \vec{\lambda}_1 -  ( V \partial_\psi - K_3 \partial_v ) \vec{\lambda}_1, \\
\vec{\mathcal{D}} ( \partial_\psi K_2 + \partial_v L_1 ) &= \vec{\mathcal{D}} \times \vec{\lambda}_2 -  ( V \partial_\psi - K_3 \partial_v ) \vec{\lambda}_2.
\label{lambdaeqn}
\end{split}
\end{equation}
The remainder of the equations can be organized (after some manipulation) into pairs that exhibit manifest symmetry under spectral interchange.  The first layer are given by
\begin{align}
\mathcal{L} K_1 &= - V \vec{\mathcal{D}} \cdot \vec{\lambda}_1 - 2 \, \vec{\nabla}V \cdot \vec{\lambda}_1 + V \, ( V \partial_\psi - K_3 \partial_v ) ( \partial_\psi K_1 + \partial_v L_2 ), \\
\mathcal{L} L_2 &= K_3 \vec{\mathcal{D}} \cdot \vec{\lambda}_1 + 2 \, \vec{\nabla}K_3 \cdot \vec{\lambda}_1 - K_3 \, ( V \partial_\psi - K_3 \partial_v ) ( \partial_\psi K_1 + \partial_v L_2 ),
\end{align}
and a similar pair under exchanging the subscripts $(1 \leftrightarrow 2)$.  The second layer becomes
\begin{equation}
\begin{split}
\mathcal{L} L_3 &= -2 \partial_v \big( \vec{\mathcal{D}} \cdot \vec{\omega} + \Phi \big) \\
& \qquad + 2 V \, \vec{\lambda}_1 \cdot \vec{\lambda}_2 + K_2 \, \vec{\mathcal{D}} \cdot \vec{\lambda}_1 + 2 \, \vec{\mathcal{D}} K_2 \cdot \vec{\lambda}_1 + K_1 \, \vec{\mathcal{D}} \cdot \vec{\lambda}_2 + 2 \, \vec{\mathcal{D}} K_1 \cdot \vec{\lambda}_2 \\
& \qquad - K_2 ( V \partial_\psi - K_3 \partial_v ) ( \partial_\psi K_1 + \partial_v L_2 ) - K_1 ( V \partial_\psi - K_3 \partial_v ) ( \partial_\psi K_2 + \partial_v L_1 ) \\
& \qquad - 2 \, ( V \partial_\psi - K_3 \partial_v ) K_2 \, ( \partial_\psi K_1 + \partial_v L_2 ) - 2 \, ( V \partial_\psi - K_3 \partial_v ) K_1 \, ( \partial_\psi K_2 + \partial_v L_1 ) \\
& \qquad + 2 V \, ( \partial_\psi K_1 + \partial_v L_2 ) ( \partial_\psi K_2 + \partial_v L_1 ),
\end{split}
\end{equation}
and
\begin{equation}
\begin{split}
\mathcal{L} M &= \partial_\psi \big( \vec{\mathcal{D}} \cdot \vec{\omega} + \Phi \big) \\
& \qquad - K_3 \, \vec{\lambda}_1 \cdot \vec{\lambda}_2 + \tfrac12 L_1 \, \vec{\mathcal{D}} \cdot \vec{\lambda}_1 + \vec{\mathcal{D}} L_1 \cdot \vec{\lambda}_1 + \tfrac12 L_2 \, \vec{\mathcal{D}} \cdot \vec{\lambda}_2 + \vec{\mathcal{D}} L_2 \cdot \vec{\lambda}_2 \\
& \qquad - \tfrac12 L_1 ( V \partial_\psi - K_3 \partial_v ) ( \partial_\psi K_1 + \partial_v L_2 ) - \tfrac12 L_2 ( V \partial_\psi - K_3 \partial_v ) ( \partial_\psi K_2 + \partial_v L_1 ) \\
& \qquad - ( V \partial_\psi - K_3 \partial_v ) L_1 \, ( \partial_\psi K_1 + \partial_v L_2 ) -  ( V \partial_\psi - K_3 \partial_v ) L_2 \, ( \partial_\psi K_2 + \partial_v L_1 ) \\
& \qquad - K_3 \, ( \partial_\psi K_1 + \partial_v L_2 ) ( \partial_\psi K_2 + \partial_v L_1 ),
\end{split}
\end{equation}
which show the spectral interchange symmetry and the dependence on $\partial_\psi K_1 + \partial_v L_2$ and $\partial_\psi K_2 + \partial_v L_1$.  Again, $\Phi$ is defined as in \eqref{Phidef}.  Finally, for $\vec{\omega}$, we have
\begin{equation}
\begin{split}
\vec {\mathcal{D}} \times \vec \omega + (V \partial_\psi - K_3  \partial_v) \vec \omega &= V \vec {\mathcal{D}} M - M\vec {\mathcal{D}} V +\frac{1}{2} \, \sum_{I=1}^3 \big( K^{I} \vec {\mathcal{D}} L_{I} - L_{I}   \vec {\mathcal{D}} K^{I} \big) \\
& \qquad - (K_2 K_3 + V L_1) \, \vec{\lambda}_1 - (K_1 K_3 + V L_2) \, \vec{\lambda}_2,
\end{split}
\end{equation}
which is invariant under spectral interchange.

%
%



\end{document}